\documentclass[pra,twocolumn,showpacs]{revtex4}
\usepackage{graphicx,graphics,psfrag,amsmath,calc,amssymb}

\begin{document}
\title{Matter-wave interference in $s$-wave and $p$-wave Fermi condensates}
\author{Wei Zhang}
\altaffiliation{Current address: Department of Physics, University of Michigan, Ann Arbor, Michigan 48109.}
\author{C. A. R. S\'{a} de Melo}
\affiliation{School of Physics, Georgia Institute of Technology,
Atlanta, Georgia 30332}

\date{\today}

\begin{abstract}
We discuss the time evolution and matter-wave interference of Fermi condensates
on the BEC side of Feshbach resonances for $s$ and $p$-wave superfluids,
upon release from harmonic traps.
In $s$-wave systems, where the order parameter is a complex scalar,
we find that the interference patterns
depend on the relative phase of the order parameters of the condensates. In $p$-wave systems
involving the mixture of two-hyperfine states,
we show that the interference pattern exhibits a polarization effect depending
on the relative orientation of the two vector order parameters.
Lastly, we also point out that $p$-wave Fermi condensates exhibit
an anisotropic expansion, reflecting the spatial anisotropy of the underlying
interaction between fermions and the orbital nature of the vector order parameter.
Potential applications of our results include systems of ultra-cold atoms
that exhibit $p$-wave Feshbach resonances such as $^6$Li or $^{40}$K.

\end{abstract}

\maketitle

\section{introduction}
\label{sec:introduction}

Matter-wave interference is a very powerful tool to study quantum
phase coherence between atomic Bose Einstein condensates
(BEC)~\cite{andrews-97,shin-04,schumm-05}, and spatial quantum noise of
bosons in optical lattices~\cite{bloch-05}. Similar techniques can
also be applied to study Fermi
condensates~\cite{greiner-03,zwierlein-03,bartenstein-04,bourdel-04,thomas-04, partridge-05},
where superfluidity can be tuned from the BCS to the BEC regime.
These experiments may reveal that the time dynamics in the BCS regime is overdamped (large Cooper
pairs can decay into two atoms), while in the BEC regime it is essentially undamped
(tightly bound molecules are stable)~\cite{sademelo-93,iskin-06a}.
Matter-wave interference experiments of $s$-wave Fermi condensates may be readily performed,
since stable condensates already exist. For $s$-wave Fermi condensates in the BEC regime quantum
interference effects are expected to be similar to those of atomic Bose condensates,
and the interference pattern should depend essentially
on the phase difference of the order parameters between
two interfering clouds.

In contrast, it is more interesting to study interference effects in $p$-wave superfluids
because of the vector nature of the order parameter. Many groups have reported some progress
towards the formation of $p$-wave Fermi condensates in harmonically trapped
clouds~\cite{regal-03b,ticknor-04,zhang-04,schunck-05} and in optical
lattices~\cite{gunter-05}, where $p$-wave Feshbach resonances have been observed.
Similar to the $s$-wave case, the scattering cross section for $p$-wave
collisions has a peak that rises over three orders of magnitude above
the small background cross section near $p$-wave resonances,
thus suggesting a divergence of the scattering parameter (volume).
Therefore, matter-wave interference can be potentially observed using Feshbach resonance
techniques when two-body dipolar or three-body losses are not too large~\cite{gunter-05}.

In addition to the diverging scattering parameter (volume), $p$-wave Feshbach resonances
also reveal characteristic features in contrast to $s$-wave counterparts.
These features include splitting of resonance peaks depending on hyperfine (pseudospin)
states as in $^6$Li and $^{40}$K (i.e., $\vert 11\rangle$, $\vert 12 \rangle + \vert 21 \rangle$,
and $\vert 22 \rangle$)~\cite{regal-03b,ticknor-04,zhang-04,schunck-05,gunter-05},
or splitting depending on angular momentum projections as in $^{40}$K
(i.e., the magnetic quantum number $m_\ell =0$ or $\pm 1$)~\cite{ticknor-04,gunter-05}.
These experimentally observed splitting allow the possibility of tuning separately $p$-wave
scattering parameters in different pseudospin and/or $m_\ell$ states,
such that the $p$-wave interaction can be in general anisotropic in both pseudospin and angular
momentum states. These additional degrees of freedom are responsible for the much
richer matter wave interference phenomena in $p$-wave Fermi condensates
than in the corresponding $s$-wave case.

Prior and in parallel to experiments the BCS to BEC evolution of $p$-wave
superfluids was discussed in the context of
ultra-cold atoms~\cite{botelho-04,botelho-05,gurarie-05,yip-05,ho-05,ohashi-05,iskin-06b},
where the existence of quantum phase transitions was
emphasized~\cite{botelho-04,botelho-05,gurarie-05,yip-05,iskin-06b}.
However, these works dealt only with zero and finite temperature
thermodynamic properties.

In this manuscript, we discuss the time evolution and matter-wave interference
of $s$-wave and $p$-wave Fermi condensates on the BEC side of Feshbach resonances,
and extend our previous work in this area~\cite{zhang-06}.
Our main results are as follows. While in atomic BEC and $s$-wave Fermi superfluids
quantum interference patterns depend essentially on the relative phase of the two clouds,
we find that in $p$-wave Fermi superfluids there can also be
a strong dependence on the relative angle
between the two vector order parameters, thus producing a polarization effect.
This polarization effect is a direct consequence
of macroscopic quantum coherence of a large number of molecules,
as well as the vector nature of the order parameter.
We also discuss the Josephson effect between two $p$-wave condensates and show
that it depends not only on the existence of phase coherence, but also on
the relative orientation of the vector order parameter of the two clouds.
Furthermore, we also show that $p$-wave Fermi condensates exhibit
an anisotropic expansion, reflecting the spatial anisotropy of the underlying
interaction between fermions and the orbital nature of the vector order parameter.

The remainder of this manuscript is organized as follows.
In Section~\ref{sec:equation-of-motion}, we discuss the dynamics of
Fermi condensates with $s$ and $p$-wave interactions.
In particular, we derive the equation of motion for the vector
order parameter of $p$-wave condensates in the strongly interacting BEC limit.
This equation of motion is further discussed in Section~\ref{sec:time-evolution},
where the time-of-flight expansion of a harmonically trapped cloud is studied.
In Section~\ref{sec:interference}, we describe matter-wave interference of
two condensates, and demonstrate that the interference pattern for the $p$-wave case
depends crucially on the relative orientation of the two vector order parameters.
In Section~\ref{sec:expansion}, we relax the restriction of being near the BEC limit
by moving towards unitarity and demonstrate that
the time-of-flight expansion of a $p$-wave Fermi condensate is in general anisotropic.
Lastly, we summarize our main results in section~\ref{sec:conclusions}.

%%%%%%%%%%%%%%%%%%%%%%%
%%%%%%%%%%%%%%%%%%%%%%%

\section{Effective theory and equation of motion}
\label{sec:equation-of-motion}

We consider a system of fermions with mass $m$ in two hyperfine
states (pseudospins), labeled by greek indices $\alpha = 1,2$. The
Hamiltonian density is (with $\hbar = k_B =1$)
\begin{eqnarray}
\label{eqn:hamiltonian}
&&{\cal H} ({\bf r}, t) = \psi_{\alpha}^\dagger({\bf r},t)
\left[-\frac{\nabla_{\bf r}^2}{2m} + U_{\rm ext}({\bf r}, t) \right]
\psi_{\alpha}({\bf r},t)
\nonumber \\
&&
- \int d{\bf r}' \Big[ \psi_\alpha^\dagger({\bf r},t)
\psi_\beta^\dagger({\bf r}',t) V_{\alpha\beta\gamma\delta} %({\bf r}-{\bf r}')
\psi_\gamma({\bf r}',t) \psi_\delta({\bf r},t) \Big],
\end{eqnarray}
where repeated greek indices indicate summation,
$\psi_\alpha^\dagger$ ($\psi_\alpha$) are creation (annihilation)
operators of fermions in state $\alpha$,
$U_{\rm ext}({\bf r}, t)$ is the time dependent trapping potential,
and $V_{\alpha\beta\gamma\delta} =  V_{\alpha\beta\gamma\delta} ({\bf r}-{\bf r}')$.
The Hamiltonian of the system is then $H (t) = \int d {\bf r} {\cal H} ({\bf r}, t)$.

The generating functional for non-equilibrium processes associated with $H(t)$ is~\cite{tokatly-04}
\begin{equation}
\label{eqn:partition-general}
Z(t) = {\rm Tr}  \hat{\cal U}^\dagger (t, t_0)
\exp\left[-\beta(H(t_0)- \mu_\alpha N_\alpha)\right]
\hat{\cal U}(t,t_0),
\end{equation}
where $N_\alpha$ is the number operator for fermions of type $\alpha$,
$\mu_\alpha$ is the corresponding chemical potential,
and $\beta = 1/T$ is the inverse temperature.
Here, $\hat{\cal U}(t,t_0) \equiv \exp[-i\int_{t_0}^t H(t')dt']$
is the time evolution operator.
This expression implicitly implies that the trapping potential
$U_{\rm ext}$ is time independent for $t<0$,
such that the initial condition corresponds to a thermal equilibrium state.
This assumption is directly related to experiments where the traps
are effectively static before the cloud release.
Therefore, the generating functional at any time $t_0<0$
takes the form $Z(t_0) = {\rm Tr} \exp\left[-\beta(H(t_0)- \mu_\alpha N_\alpha)\right]$,
which is just the partition function describing thermal equilibrium properties.

\begin{figure}
\begin{center}
\psfrag{Retau}{Re($\tau$)}
\psfrag{Imtau}{Im($\tau$)}
\psfrag{-b}{$-\beta$}
\psfrag{t}{$t$}
\psfrag{t0}{$t_0$}
\psfrag{C}{$C$}
\psfrag{0}{$0$}
\includegraphics[width=7.5cm]{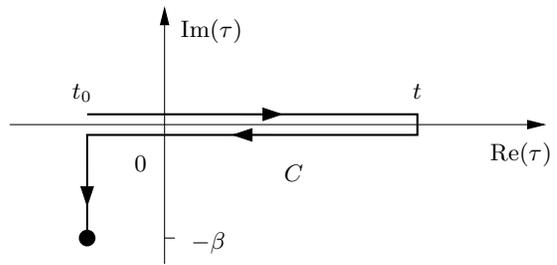}
\end{center}
\caption{Integration contour $C$ used in Eq.~(\ref{eqn:generating}).}
\label{fig:contour}
\end{figure}

By introducing a complex time $\tau$, the generating functional
Eq.~(\ref{eqn:partition-general}) can be written as
\begin{equation}
\label{eqn:generating}
Z(t)= \int_{\rm BC} {\cal D}[\psi_\alpha^\dagger, \psi_\alpha]
e^{\left\{ - S_2[\psi_\alpha^\dagger, \psi_\alpha]
- S_4[\psi_\alpha^\dagger, \psi_\alpha] \right\}},
\end{equation}
where the boundary condition (BC) is antisymmetric over the
integration contour $C$ of the functional integral
(as shown in Fig.~\ref{fig:contour}),
and $S_2$ and $S_4$ are quadratic and quartic action functions
of fermionic operators $\psi$, respectively.

In what follows, we discuss the $p$-wave case in detail and quote
the more standard results for $s$-wave. To calculate the effective
action of the system, we write $V_{\alpha\beta\gamma\delta}(\mbox{\boldmath$\rho$}) =
V(\mbox{\boldmath$\rho$}) \Gamma_{\alpha\beta\gamma\delta}$,
where in a triplet channel $\Gamma_{\alpha\beta\gamma\delta}=
{\bf v}_{\alpha\beta} \cdot ({\bf v}^\dagger)_{\gamma\delta}$.
The pseudospin matrix elements for the $j^{\rm th}$ component of ${\bf v}$ are
$(v_j)_{\alpha\beta} \equiv (i \sigma_j \sigma_y)_{\alpha\beta}$,
where $\sigma_j$ are Pauli matrices.
By writing the interaction in this form, we implicitly assume that
the interaction is symmetric in pseudospin space
with $V_{1111}=V_{1212}=V_{2222}$.
It should be emphasized that this assumption is introduced
just for simplicity and the resulting equation of motion is valid for a more
general interaction. In fact, the experimental realizations of the polarization
effect in mater-wave interference and the anisotropic expansion proposed here rely
on the tunability of the interaction in both pseudospin and angular momentum ($m_\ell$) spaces.
Here, we first consider the simpler case of symmetric interaction,
and postpone the general discussion until later in Section~\ref{sec:interference}.

In order to integrate out the fermions, we introduce the field
${\bf B}^\dagger({\bf r}, {\bf r}' ,\tau) =
\psi_\alpha^\dagger({\bf r},\tau) {\bf v}_{\alpha\beta} \psi_\beta^\dagger({\bf r}',\tau)$
and the corresponding auxiliary field ${\bf d}({\bf r}, {\bf r}', \tau)$.
Since the ${\bf d}$ field depends on two spatial variables ${\bf r}$ and ${\bf r}^\prime$,
it can also be transformed into the center-of-mass ${\bf R}=({\bf r}+{\bf r}')/2$
and relative $\mbox{\boldmath$\rho$}={\bf r}-{\bf r}'$ coordinates:
\begin{equation}
\label{eqn:dvector}
{\bf d}({\bf r}, {\bf r}', \tau) =
\sum_{n, \ell, m_{\ell}} {\bf D}_{n,\ell, m_{\ell}} ({\bf R},\tau)
\eta_{n, \ell, m_{\ell}} (\mbox{\boldmath$\rho$}),
\end{equation}
where $\eta_{n,\ell, m_{\ell}}(\mbox{\boldmath$\rho$})$ are eigenfunctions
of the reduced two-body Hamiltonian
${\cal H}_2 = -\nabla_{\rho}^2/m + V(\mbox{\boldmath$\rho$})$.

The vector nature of the order parameter ${\bf d}$ or ${\bf D}$
can be understood from the pseudospin structure of the pair wave function
$\Psi_{\rm pair} = g_{11} \vert 11 \rangle
+ g_{12} (\vert 12 \rangle + \vert 21 \rangle) + g_{22} \vert 22 \rangle$.
Using the symmetric matrices $i\sigma_j \sigma_y$,
this pair wave function can be represented in the form
$ \Psi_{\rm pair} = i {\bf d} \cdot \mbox{\boldmath$\sigma$} \sigma_y$,
where the components of ${\bf d}$ are related to the amplitudes $g_{ij}$ through
$g_{11} = -d_x + i d_y$, $g_{12} = d_z$, and $g_{22} = d_x + i d_y$.
Therefore, the direction of ${\bf d}$ (or ${\bf D}$)
determines the amplitude of the pair wavefunction in each of the pseudospin triplet channels
$\vert 11 \rangle$, $\vert 12 \rangle + \vert 21 \rangle$, and $\vert 22 \rangle$.
For $^6$Li these orthogonal states correspond to the $\nu = 38$ vibrational
state of the singlet potential~\cite{schunck-05} of total electronic spin $S = 0$,
and total nuclear spin $I = 1$.

For definiteness, we consider a pure $p$-wave interaction
where the ground state is three-fold degenerate ($\ell = 1$)
and labeled by $m_\ell = -1, 0,1$.
A rotation of basis from spherical harmonics
$Y_{1,m_\ell}(\hat{\mbox{\boldmath$\rho$}})$
to $p_{\nu =x,y,z}$ with corresponding eigenfunctions
$\eta_{0, 1}(\rho) \hat{\mbox{\boldmath$\rho$}}_\nu$
leads to ${\bf d} = \sum_{\nu} {\bf D}_{\nu} ({\bf R},\tau)
\eta_{0,1}(\rho) \hat{\mbox{\boldmath$\rho$}}_\nu$
at low temperatures, where the higher energy states are not excited.
In the BEC limit,
Cooper pairs are tightly bound molecules and the relative degrees
of freedom \mbox{\boldmath$\rho$} can be integrated out leading to
an effective action
\begin{eqnarray}
\label{eqn:effaction}
S_{\rm eff} &=& - \int d\tau \int d{\bf R}
\Big \{
{\bf D}^\dagger({\bf R},\tau) \cdot \left[\hat{\cal K}{\bf D} ({\bf R},\tau)\right]
\nonumber \\
&& - \frac{g_0}{2} \left[ 2 \vert {\bf D} ({\bf R},\tau)\vert^4 -
\vert {\bf D}^2({\bf R},\tau)\vert^2 \right] \Big\},
\end{eqnarray}
where $D_j ({\bf R}) \equiv \sum_{\nu} D_{j,\nu}({\bf R})$,
and the operator $\hat{\cal K} = i\partial_\tau - 2 U_{\rm ext}({\bf R},\tau)
+ \nabla_{\bf R}^2 / (4m)$ corresponds to the action of an ideal
non-equilibrium gas of Bose particles with mass $M=2m$. This action leads to equations of motion
\begin{eqnarray}
\label{eqn:GP}
i \partial_t D_j &=& \left[ -\frac{\nabla_{\bf R}^2}{2M} + 2 U_{\rm ext} ({\bf R},t)
+ 2 g_0 \vert {\bf D} \vert^2 \right] D_j
\nonumber \\
&&
\hspace{1cm}
- g_0 \left({\bf D}\cdot {\bf D} \right) D_j^\dagger.
\end{eqnarray}
Notice that this expression is different from the time-dependent
Gross-Pitaevskii (TDGP) equation for an atomic vector boson field.
The difference comes from the last term, which describes a non-unitary
complex order parameter of the underlying paired fermions.
In contrast, the standard TDGP equation for scalar atomic bosons is obtained
in the $s$-wave case~\cite{tokatly-04}.

Equation (\ref{eqn:GP}) can be simplified to the TDGP form in two special cases.
First, if the atomic hyperfine states $\vert 1\rangle$ and $\vert 2 \rangle$
are equally populated with $N_1 = N_2 = N$ ($\mu_1 = \mu_2 = \mu$) and ${\bf D}$ is unitary,
then  ${\bf D}$ is a real vector with an overall phase,
leading to the equation of motion
\begin{equation}
\label{eqn:GPunitary}
i\partial_t D_j = \left[ -\frac{\nabla_{\bf R}^2}{2M} + 2 U_{\rm ext} ({\bf R},t)
+ g_0 \vert {\bf D} \vert^2 \right] D_j.
\end{equation}
Second, if only one atomic hyperfine state is populated,
then ${\bf D}$ is non-unitary and ${\bf D} = A (1, \pm i, 0)$,
where $A$ is a complex constant.
Thus, the last term ${\bf D}\cdot {\bf D}$ in Eq.~(\ref{eqn:GP}) vanishes,
and the equation of motion is identical to Eq.~(\ref{eqn:GPunitary}),
with $g_0 \to 2g_0$. In the following section,
we confine ourselves to these two special cases
and discuss the time-of-flight expansion of a
Fermi condensate upon released from the trap.

%%%%%%%%%%%%%%%%%%%
%%%%%%%%%%%%%%%%%%%
\section{Time-of-flight expansion of a harmonically trapped cloud}
\label{sec:time-evolution}

In order to describe the time evolution of a triplet Fermi condensate,
one needs to solve the nonlinear equation of motion Eq. (\ref{eqn:GPunitary}),
which usually requires special numerical treatment.
For simplicity and definiteness, we consider a Fermi superfluid
is released from a harmonic trap, i.e.,
\begin{equation}
\label{eqn:trap}
U_{\rm ext}({\bf r},t) = \sum_{j=x,y,z} m\omega_j^2(t) r_j^2/2,
\end{equation}
where $\omega_j(t<0) = \omega_{j}$ are constants
and $\omega_j(t\ge 0) = 0$.
Thus, for $t<0$, the system is described by
\begin{equation}
\label{eqn:staticGP}
\mu_0 D_j({\bf R}) =
\left[ -\frac{\nabla_{\bf R}^2}{2M} + 2 U_{\rm ext} ({\bf R})
+ g \vert {\bf D} \vert^2 \right] D_j,
\end{equation}
where $\mu_0$ is the effective boson chemical potential, and $g = g_0$ ($g = 2g_0$)
when ${\bf D}$ is unitary (non-unitary). For dominant Boson interactions
the Thomas-Fermi approximation leads to
\begin{equation}
\label{eqn:thomas-fermi}
\vert {\bf D}({\bf R},0) \vert =
\left\{\begin{array}{l}
\sqrt{\frac{\mu_0 - 2U_{\rm ext}({\bf R})}{g}},
\textrm{ for } \mu_0 \ge 2U_{\rm ext}({\bf R})
\\
0, \textrm{ otherwise.}
\end{array} \right.
\end{equation}
When this approximation fails the initial condition for the time evolution can be obtained by
solving Eq.~$(\ref{eqn:staticGP})$ numerically.

For $t >0$, we use the transformation $ \label{eqn:scale}R_j(t) = b_j(t) R_j(0)$,
where the scaling factors $b_j(t)$ satisfy~\cite{kagan-96,castin-96},
\begin{equation}
\label{eqn:bj} \frac{d^2 b_j(t)}{dt^2} =
\frac{\omega_{j}^2}{A(t) b_j(t)}
\end{equation}
with $A(t) = b_x(t) b_y(t) b_z(t)$ and initial conditions $b_j(0)= 1$.
For a cigar-shaped trapping potential with axial symmetry
($\omega_{x} = \omega_{y} \equiv \omega_{\perp} \gg \omega_{z}$),
$D_j({\bf R},t)$ becomes
\begin{equation}
\label{eqn:Dlimit}
D_j({\bf R},t) \approx \frac{\exp[iS({\bf R},t)]}{\sqrt{1+ \lambda^2}}
D_j(\overline{\bf R},0),
\end{equation}
where $\overline{R}_k = R_k/b_k(t)$ are scaled coordinates,
$\lambda \equiv \omega_\perp t$ is the dimensionless time,
and the phase factor
\begin{equation}
\label{eqn:phase}
S({\bf R}(t), t) = S_0(t) + M \sum_{k} \frac{R_k^2(t) }{b_k(t)} \frac{d b_k(t)}{dt}.
\end{equation}
The result for $s$-wave is formally identical to that of
Eq.~(\ref{eqn:Dlimit}) with the substitution $D_j \to \Psi$, where
$\Psi$ represents the scalar order parameter.

In the limit where $\varepsilon \equiv \omega_z/\omega_\perp \ll 1$,
approximated solutions for $b_j(t)$ can be obtained as a power
expansion of $\varepsilon$, leading to
\begin{eqnarray}
\label{eqn:phase2}
S({\bf R}(t), t) &=& - \frac{\mu_0 \tan^{-1}(\lambda)}{\omega_\perp \lambda}
+ \frac{\mu_0 \varepsilon^2}{\omega_\perp} \theta(\lambda)
\nonumber \\
&&
\hspace{-20mm}
+ M \frac{\omega_\perp \lambda}{1+\lambda^2} (R_x^2 + R_y^2)
+ M \varepsilon^2 \omega_\perp \tan^{-1}(\lambda) R_z^2,
\end{eqnarray}
where the function $\theta(\lambda)$ in the second term takes the following form
\begin{equation}
\label{eqn:theta}
\theta(\lambda) =
\int_0^\lambda \frac{2x\tan^{-1}(x) - \ln(1+x^2)}{2(1+x^2)} dx.
\end{equation}
Notice that $\theta(\lambda) \sim \ln (\lambda)$ as $t \to \infty$,
hence the second term in Eq. (\ref{eqn:phase2}) becomes comparable
to the first one when $t \ln (\lambda) \sim \varepsilon^2$.
For realistic experimental parameters where
$\varepsilon \sim 10^{-2}$ and $\omega_\perp \sim 10^3 $ Hz,
this condition is satisfied only when $t\sim 10^2$ s,
which is several orders of magnitude larger than any time scales
in current experiments~\cite{shin-04,schumm-05}.
Therefore, the $\theta(\lambda)$ term in Eq. (\ref{eqn:phase2})
can be neglected in the following discussion.

It should be emphasized that the scale transformation discussed above
is valid under the hydrodynamic approximation,
where the collisions between molecules are not playing a crucial role.
Furthermore, it is also assumed that the Cooper pairs (Feshbach molecules)
are long lived and the damping processes are negligible.
With these conditions in mind, we discuss next matter wave interference
of two clouds.

%%%%%%%%%%%%%%%%
%%%%%%%%%%%%%%%%
\section{matter-wave interference and polarization effect}
\label{sec:interference}

In this section, we consider first the matter-wave interference of two spatially separated condensates
such that the energy barrier between them is large enough to neglect the tunneling effect.
Thus, one may write down the total wave function as
\begin{equation}
\label{eqn:coherentstate}
\Phi_{\rm tot}({\bf R},t) = \Phi_{\rm
L}({\bf R},t) + \Phi_{\rm R} ({\bf R},t),
\end{equation}
where $\Phi_{\rm P}\propto i \sum_j D_{j,{\rm P}} \sigma_j \sigma_y$
denotes the pair wavefunction of Fermi condensate
in the left (${\rm P}={\rm L}$) or right (${\rm P}={\rm R}$) trap.
The left and right trap centers lie at $( -W/2, 0,0)$ and $( W/2, 0,0)$,
respectively, where $W$ is the distance between traps.
We consider the case of two identical axially symmetric Fermi condensates
with $\omega_\perp \gg \omega_z$.
In this case, the time evolution of each cloud is described by
\begin{equation}
\label{eqn:Dlt}
D_{j,{\rm P}} ({\bf R},t) = \frac{\exp[iS({\bf R}\pm W\hat{\bf x}/2,t)]}{\sqrt{1+ \lambda^2}}
D_{j,{\rm P}} (\overline{{\bf R}\pm W\hat{\bf x}/2},0),
\nonumber
\end{equation}
Thus, for a single run of experiment, the particle density
$n({\bf R},t)\equiv \vert \Phi_{\rm tot}({\bf R},t) \vert^2$ is
\begin{eqnarray}
\label{eqn:n}
n({\bf R}, t) &\propto& \vert {\bf D}_{\rm L}({\bf R},t) \vert^2
+ \vert {\bf D}_{\rm R}({\bf R},t) \vert^2
\nonumber \\
&& \hspace{-1.8cm}
+ 2 {\rm Re}
\frac{{\bf D}_{\rm L}^\dagger (\overline{{\bf R}+W\hat{\bf x}/2},0)
\cdot {\bf D}_{\rm R}(\overline{{\bf R}-W\hat{\bf x}/2},0) e^{i\chi}}
{A(\lambda)},
\end{eqnarray}
where the time dependent phase factor
$\chi({\bf R},t) = S({\bf R}+W\hat{\bf x}/2,t )- S({\bf R}-W\hat{\bf x}/2,t )+ \chi_0$,
and $\chi_0$ is the initial relative phase of the two condensates.
The result for $s$-wave is formally identical to that of Eq.~(\ref{eqn:n})
with the substitution $D_j \to \Psi$.

\begin{figure}
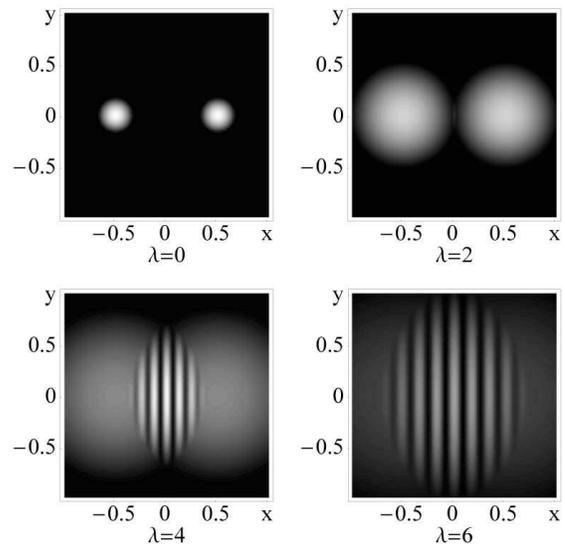

\begin{center}
\includegraphics[width=3.5cm]{fig2a_s.eps}
\hspace{1mm}
\includegraphics[width=3.5cm]{fig2b_s.eps}
\vskip 2mm
\includegraphics[width=3.5cm]{fig2c_s.eps}
\hspace{1mm}
\includegraphics[width=3.5cm]{fig2d_s.eps}
\end{center}
\caption{Interference pattern versus dimensionless time $\lambda = \omega_\perp t$
for $p$-wave Fermi condensates in the BEC limit with $\omega_z/\omega_\perp=0.1$,
assuming $\vert {\bf D}_{\rm L}^{\dagger} \cdot {\bf D}_{\rm R}\vert$ is maximal.
The plots include only the superfluid part, and show columnar density
versus $x,y$ coordinates in units of the initial clouds separation $W$.
The patterns are similar to those of atomic scalar bosons, and $s$-wave paired Fermions.
}
\label{fig:inter}
\end{figure}

\begin{figure}
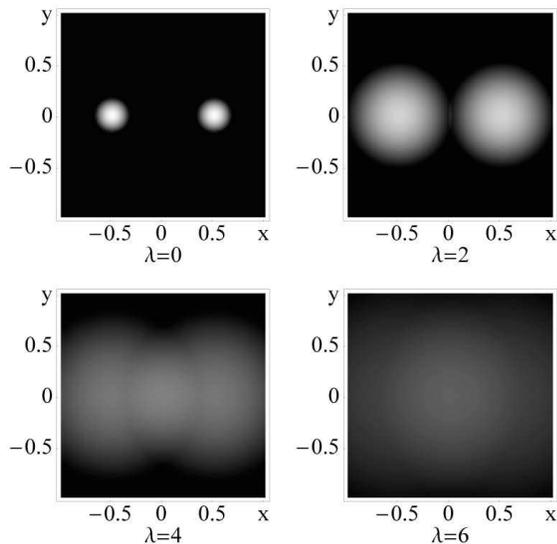

\begin{center}
\includegraphics[width=3.5cm]{fig2a_s.eps}
\hspace{1mm}
\includegraphics[width=3.5cm]{fig2b_s.eps}
\vskip 2mm
\includegraphics[width=3.5cm]{fig3c_s.eps}
\hspace{1mm}
\includegraphics[width=3.5cm]{fig3d_s.eps}
\end{center}
\caption{Interference pattern versus dimensionless time $\lambda = \omega_\perp t$
for $p$-wave Fermi condensates in the BEC limit with $\omega_z/\omega_\perp=0.1$,
assuming $\vert {\bf D}_{\rm L}^{\dagger} \cdot {\bf D}_{\rm R}\vert = 0$.
The plots include only the superfluid part, and show columnar density versus
$x, y$ coordinates in units of the initial clouds separation $W$.
}
\label{fig:nointer}
\end{figure}

When each cloud has the same hyperfine state occupied (e.g. $\Psi_{\rm pair} = g_{11} \vert 11 \rangle$)
the ${\bf D}$ vectors in each cloud have the fixed form $A(1,i,0)$
and fringes are present in all experimental realizations.
This result is similar to the $s$-wave case where the order parameter is a complex scalar.
However, when both Fermi condensates are in unitary states,
${\bf D}$ is essentially a real vector with an overall phase,
and $n({\bf R},t)$ shows an angular dependence controlled by the dot product term in Eq.~(\ref{eqn:n}).
When the two order parameters are parallel, this term is maximal and the interference pattern is
most visible (Fig.~\ref{fig:inter}).
But if the ${\bf D}$ vectors are perpendicular,
fringes are absent at all times (Fig.~\ref{fig:nointer}).
Therefore, in the unitary case the existence and intensity of
interference fringes are very sensitive to the relative
orientation of the ${\bf D}$ vectors.

This sensitivity to the relative orientation of the order parameters and
the corresponding polarization effect also manifest themselves in the
Josephson tunneling between two condensates. By considering a tunneling process across the energy barrier,
the left and right condensates can be described by the modified equations of motion
\begin{subequations}
\label{eqn:TDGP-tunnel}
\begin{eqnarray}
i\partial_t D_{{\rm L},j} &=&
\left[ -\frac{\nabla_{\bf R}^2}{2M} + 2 U_{\rm ext} ({\bf R})
+g \vert {\bf D}_{\rm L} \vert^2 \right] D_{{\rm L},j}
\nonumber \\
&& \hspace{-5mm}
+ \sum_k \int d{\bf R}^\prime T_{kj}({\bf R}, {\bf R}^\prime)
D_{{\rm R},j}({\bf R}^\prime,t),
\\
i\partial_t D_{{\rm R},j} &=&
\left[ -\frac{\nabla_{\bf R}^2}{2M} + 2 U_{\rm ext} ({\bf R})
+g \vert {\bf D}_{\rm R} \vert^2 \right] D_{{\rm R},j}
\nonumber \\
&& \hspace{-5mm}
+ \sum_k \int d{\bf R}^\prime T_{jk}^* ({\bf R}, {\bf R}^\prime)
D_{{\rm L},j}({\bf R}^\prime,t),
\end{eqnarray}
\end{subequations}
where $T_{jk}({\bf R}, {\bf R}') \equiv T_c + T_s$ is the tunneling matrix,
with $T_c$ and $T_s$ are the pseudospin-conserving and non-conserving portions, respectively.
In the case where the trapping potential is much smaller than the
energy difference between the two hyperfine states,
or the time scale of experiments is much shorter than the lifetime
of each hyperfine state, the non-conserving tunneling process is negligible,
leading to $T_{jk} = T({\bf R}, {\bf R}') \delta_{jk}$.
Thus, the Josephson current $J$ is dominated by the pseudo-spin-conserving tunneling
processes, leading to
\begin{equation}
\label{eqn:current}
J ={\rm Im} \int_{LS} d{\bf R} \int d{\bf R}' T({\bf R}, {\bf R}')
{\bf D}_{\rm L}^\dagger ({\bf R}, t) \cdot {\bf D}_{\rm R}({\bf R}', t).
\end{equation}
Notice that the Josephson current depends crucially on the relative
orientation of the two vector order parameters,
and acquires a polarization effect as in matter-wave interference.
In particular, if ${\bf D}_{\rm L}^\dagger ({\bf R}) \cdot {\bf D}_{\rm R}({\bf R}') = 0$
is satisfied, there is no Josephson tunneling between the two condensates
for all ${\bf R}$ and ${\bf R}'$. This property is similar to that encountered
in Josephson junctions of lattice $p$-wave superconductors, where the Josephson
tunneling current between two $p$-wave crystals depend on the relative orientation
of the vector order parameters~\cite{vaccarella-03a, vaccarella-03b}.

Since the interference pattern and the Josephson current depend
crucially on ${\bf D}_{\rm L}$ and ${\bf D}_{\rm R}$,
it is important to understand how these vectors can be controlled experimentally.
In the discussion above we assumed a symmetric interaction in pseudospin space,
i.e., $V_{1111}$, $V_{1212}$ and $V_{2222}$ were identical.
However, experimental results for $p$-wave Feshbach resonances show a
finite separation in different channels~\cite{schunck-05} implying different
interactions in pseudospin space. When the different interaction strengths
are absorbed into an effective ${\bf D}$ vector, an equivalent procedure
leads to equations similar to Eqs. (\ref{eqn:GP}) and (\ref{eqn:GPunitary}).
For instance, the $p$-wave resonances for $^6$Li occur at $159$G (width $0.4$G),
$185$G (width $0.2$G) and $215$G (width $0.4$G)
for the $\vert 11\rangle$, $\vert 12 \rangle +\vert 21 \rangle$, and
$\vert 22 \rangle$ channels, respectively.
By applying a constant plus a gradient magnetic field,
the local field at L (R) cloud can be tuned to be $216~(214)$G,
which is above (below) the $\vert 22\rangle$ resonance.
Thus, a sweep down of the constant magnetic field
by $30$G makes the L cloud cross the $\vert 22 \rangle$
but not the $\vert 12 \rangle$ resonance, while it makes the R cloud cross only the
$\vert 12 \rangle$ resonance. In this case, the L cloud is in the BEC regime of the $\vert 22 \rangle$
channel, with $\Psi_{{\rm pair, L}} \approx g_{22,{\rm L}} \vert 22 \rangle$
or ${\bf D}_{\rm L} = g_{22,{\rm L}} (1/2, -i/2,0)$.
However, the R cloud is in the BEC regime of the $\vert 12 \rangle$ channel,
with $\Psi_{{\rm pair, R}} \approx g_{12,{\rm R}} (\vert 12 \rangle + \vert 21 \rangle)$
or ${\bf D}_{\rm R} = g_{12,{\rm R}} (0, 0, 1)$.
After these initial states are prepared the total magnetic field $B({\bf r})$ (constant + gradient)
and the harmonic trapping potential are turned off suddenly (very fast) and simultaneously,
as it is standard in experiments~\cite{zwierlein-04-prl, chin-06}, such that the magnetic field
does not interfere with the subsequent cloud expansion.
Therefore, ${\bf D}^{\dagger}_{\rm L} \cdot {\bf D}_{\rm R} = 0$
and the interference pattern is that of Fig.~\ref{fig:nointer}.
More generally, for a given magnetic field gradient
one can choose the field each cloud is subjected to by
adjusting the relative distance between the clouds, and control the
pair wavefunctions (or {\bf D} vectors).

%%%%%%%%%%%%%%%%%%%%%%%%%%%%%%
%%%%%%%%%%%%%%%%%%%%%%%%%%%%%
\section{Anisotropic expansion}
\label{sec:expansion}

Until now, we considered Fermi condensates trapped in a harmonic potential only in the
strongly interacting BEC limit, where fermions form tightly bound molecules
and the internal degrees of freedom of the fermion pairs do not play an important role.
However, if one moves away from the BEC limit towards unitarity, the average pair size
increases with decreasing interaction strength, and the internal structure of fermion pairs
can dramatically change the condensate properties when the pair size becomes comparable
to the inter-molecular spacing.

In this section, we discuss the expansion of a harmonically trapped cloud away
from the BEC limit (but still on the BEC side of the Feshbach resonance).
In such case, the method used to derive the effective action Eq. (\ref{eqn:effaction})
is not directly applicable, since the internal-degree-of-freedom wave function
is no longer localized within a small volume. However, if the trapping potential
$U_{\rm ext}$ varies slowly in comparison to the
coherence length, a semiclassical approximation can be applied.
Within this approximation, one can first derive the effective action in a free space
(without trapping potential), and add the potential afterwards. Again, we consider here only
two cases: a non-unitary case where fermions are in a single hyperfine state,
and unitary cases where fermions are in two equally populated hyperfine states.

In such cases, the quadratic term of the effective action takes the form
\begin{equation}
\label{eqn:5-S2-weak}
S_{\rm eff}^{(2)} = -\int dt \int d{\bf R}
{\bf D}_{m_\ell}^\dagger({\bf R},t) \cdot
\hat{\cal L} {\bf D}_{m_\ell}({\bf R},t),
\end{equation}
where $\hat{\cal L} =
a_{m_\ell} - \sum_{ij} c_{m_\ell}^{ij} \nabla_i \nabla_j/(4m)
+ 2 U_{\rm ext}({\bf R}) + i d_{m_\ell} \partial_t$.
Here, we use the basis of spherical harmonics $Y_{1,m_\ell}$ and define
$D_{j,m_\ell}$ by $d_j({\bf r}, {\bf r}^\prime) = \sum_{m_\ell}
D_{j,m_\ell}({\bf R}) \eta_{0,1}(\rho) Y_{1,m_\ell}(\hat{\mbox{\boldmath$\rho$}})$,
and assume that the order parameter is dominated and characterized by
the spherical harmonics with either $Y_{1,m_\ell=0}$ or $Y_{1,m_\ell=\pm1}$.
The coefficients $a$, $c^{ij}$, and $d$ in Eq. (\ref{eqn:5-S2-weak})
can be obtained by considering the free space problem and transforming it
into momentum space~\cite{iskin-06a}.
For a weak trapping potential, the coefficient $c_{m_\ell}^{ij}$ becomes
\begin{eqnarray}
\label{eqn:cij}
c_{m_\ell}^{ij} &=& \sum_{\bf k}
\Bigg\{ \left[\frac{X({\bf k})}{4 E^2({\bf k})}
- \frac{\beta Y({\bf k})}{16 E({\bf k})}\right]
\delta_{ij}
\nonumber \\
&& \hspace{1cm}
+ \kappa_{m_\ell}^{ij}
\frac{\beta^2 {\bf k}^2 X({\bf k}) Y ({\bf k})}{32 m E({\bf k})}
\Bigg\} \phi^2(k),
\end{eqnarray}
where $E({\bf k}) = \xi_{1,{\bf k}} + \xi_{2,{\bf k}}$,
$\xi_{\alpha,{\bf k}} = {\bf k}^2/2m - \mu_\alpha$,
$X({\bf k}) = \tanh(\beta \xi_{1,{\bf k}}/2) + \tanh(\beta\xi_{2,{\bf k}}/2)$,
and $Y({\bf k}) = {\rm sech}^2(\beta\xi_{1,{\bf k}}/2)
+ {\rm sech}^2(\beta\xi_{2,{\bf k}}/2)$.
The symmetry function $\phi(k)$ is defined by
$V({\bf k}, {\bf k}') = \int d{\mbox{\boldmath$\rho$}}
V({\mbox{\boldmath$\rho$}}) \exp [i({\bf k}-{\bf k}')\cdot {\mbox{\boldmath$\rho$}}]
= V \phi(k) \phi(k') Y_{1,m_\ell}(\hat{\bf k}) Y_{1,m_\ell^\prime}^*(\hat{\bf k}')$,
and the angular average is
\begin{equation}
\kappa_{m_{\ell}}^{ij} =
\int d\hat{\bf k} \hat{k}_i \hat{k}_j
Y_{1,m_\ell}(\hat{\bf k}) Y_{1,m_{\ell}}^*(\hat{\bf k})
= \kappa_{m_\ell}^{ii} \delta_{ij}.
\end{equation}

For the case where $m_\ell = 0$, this angular average is
anisotropic with $\kappa_0^{yy} = \kappa_0^{zz} = 1/10$, and $\kappa_0^{xx} = 3/5$.
Here, we choose the $x$ direction to be the quantization axis.
Therefore, the coefficient $c_0^{ij}$ (which is directly related to
the Ginzburg-Landau coherence length $\xi_{ij}$) is diagonal and anisotropic,
hence acquiring a mass anisotropy $M_{i} = 2m/c_0^{ii}$.
This mass anisotropy reflects the higher angular momentum ($p$-wave)
nature of the order parameter for paired fermions, and it is
absent for $s$-wave Fermi and atomic Bose condensates.

\begin{figure}
\begin{center}
\hspace{-3mm}
\includegraphics[width=3.5cm]{fig4a_s.eps}
\hspace{2mm}
\includegraphics[width=2.6cm]{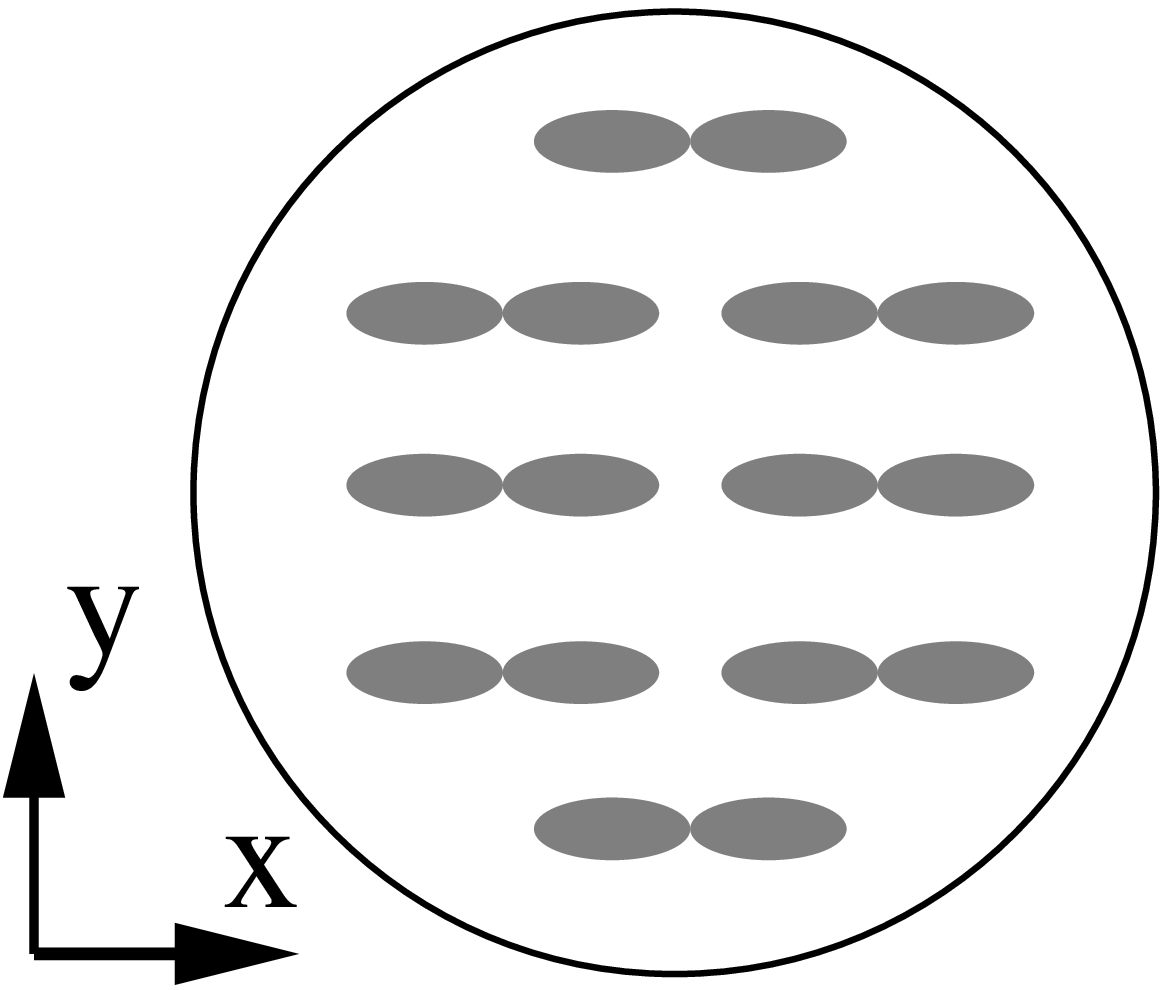}
\vskip 2mm
\includegraphics[width=3.5cm]{fig4c_s.eps}
\hspace{0mm}
\includegraphics[width=3.0cm]{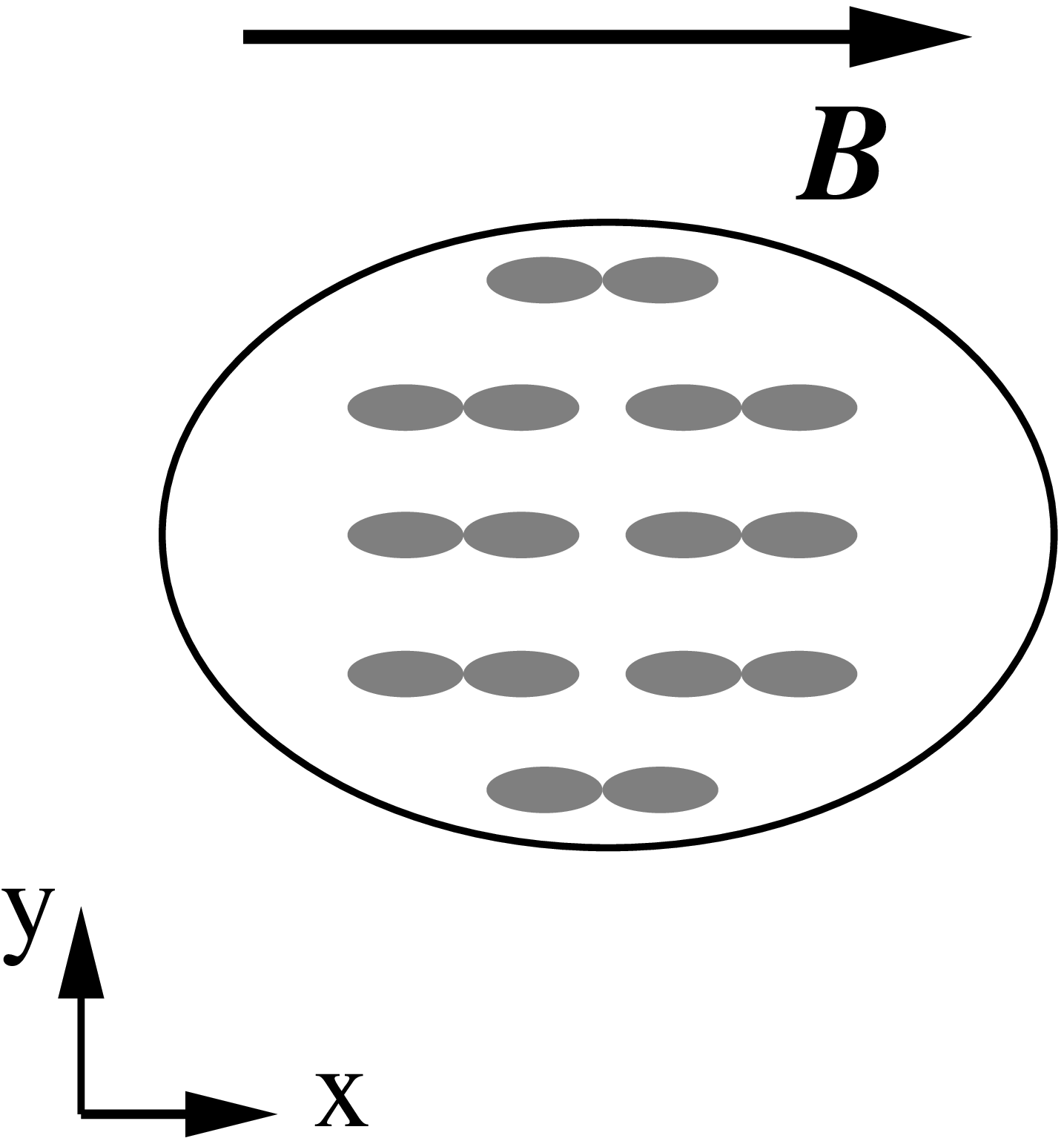}
\end{center}
\caption{
$p$-wave Fermi condensate of $p_x$-symmetry ($m_\ell=0$)
(a) in an axially symmetric trap and (b) upon release from trap.
Notice that the axial symmetry is lost in the $xy$ plane
due to the anisotropic effective mass (interaction).
}
\label{fig:aniso-expan}
\end{figure}

As an example, we consider a $p$-wave Fermi condensate
in an axially symmetric trap where $\omega_x=\omega_y \gg \omega_z$,
with a magnetic field applied along $\hat{\bf x}$ (chosen as the quantization axis)
to tune through the Feshbach resonance (see Fig.~\ref{fig:aniso-expan}).
Since the resonances for different $m_\ell$ states are split as in $^{40}$K~\cite{ticknor-04,gunter-05},
it may be possible to adjust the magnetic field such that fermions are
paired in the $m_\ell=0$ ($p_x$) state only.
In this case, the $p$-wave interaction leads to the formation
of $p_x$ symmetry pairs, which are more strongly correlated
along the $x$ direction ($\xi_x > \xi_y = \xi_z$ or $M_x < M_y = M_z$).
Thus, it is easier to accelerate the cloud along the direction of
lighter mass $M_x$ such that the cloud expands faster along the $x$-direction
than along the $y$-direction, hence breaking the axial symmetry.
This anisotropic expansion due to $p$-wave interactions also occurs
for a completely isotropic trap, and it is very different from the
anisotropy inversion (in the $xz$ and $yz$ planes) found in axially symmetric traps
for $s$-wave Fermi condensates~\cite{menotti-02}.
The anisotropy inversion is related only to the anisotropy of trapping potential,
while the anisotropic expansion discussed here is due to anisotropic interactions.

\begin{figure}
\begin{center}
\psfrag{r}{$r_L$}
\psfrag{rM}{$r_M$}
\includegraphics[width=7.5cm]{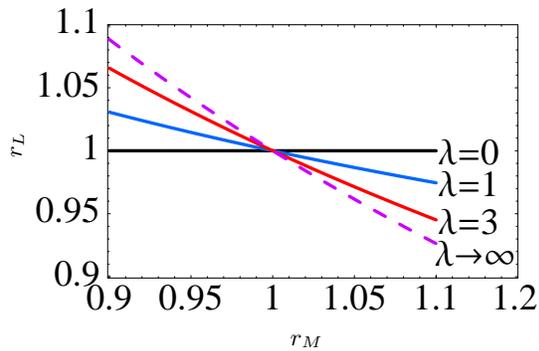}
\end{center}
\caption{Cloud anisotropy ratio $r_L = L_x/L_y$ as a function of
effective mass anisotropy ratio $r_M = M_x/M_y$ at time $\lambda$ (Solid lines).
Dashed line indicates the saturated behavior at $\lambda \to \infty$.}
\label{fig:ratio}
\end{figure}

In Fig.~\ref{fig:ratio}, we show the cloud anisotropy ratio $r_L = L_x/L_y$
as a function of the effective mass anisotropy ratio $r_M = M_y/M_x = \xi_x^2/\xi_y^2$.
The anisotropy effect disappears in the BEC limit as the effective masses become isotropic,
but becomes more evident towards unitarity.
The values of $r_M$ change as a function of the scattering volume $a_p$,
and vary from $r_M = 1$ in the BEC limit ($a_p \to 0^+$), to
$r_M = 3$ in the BCS limit ($a_p \to 0^-$). Since our theory is valid only on the
BEC side where the fermion chemical potential $\mu < 0$,
the maximal theoretical anisotropy is reached near $\mu = 0$
(which is also close to the unitarity limit $a_p \to \pm \infty$~\cite{iskin-06a}),
leading to a $10\%$ anisotropy ($r_M \approx 1.1$)
for trapped $^{40}$K in the $p_x$-state $(m_{\ell} = 0)$.
Investigations on the BCS side ($\mu > 0$) and at unitarity
require the inclusion of Landau damping which
leads to the decay of Cooper pairs, and are beyond
the scope of the present theory~\cite{naraschewski-96}.
However, even in this regime fermion pairs survive
at least initially in time-of-flight experiments for $s$-wave systems~\cite{schunck-06}.
%%
%

%%%%%%%%%%%%%%%%%%%%%%%%
%%%%%%%%%%%%%%%%%%%%%%%%
\section{conclusions}
\label{sec:conclusions}

In summary, we considered Fermi condensates
with $s$- and $p$-wave interactions (e.g. $^6$Li and $^{40}$K), and derived the
equation of motion for the $p$-wave case in a vector boson representation on the BEC side
of the Feshbach resonance.
We derived general equations of motion for the order parameter in the unitary and non-unitary
$p$-wave cases,
and showed that the equation of motion can be simplified to a similar form as
time dependent Gross-Pitaevskii (TDGP) equation for vector atomic Bose systems,
when the superfluid consists of fermions
in a single hyperfine state, or in two equally populated hyperfine states.
Within these two special cases, we described the time evolution of the vector
order parameter approximately by scaling the spatial coordinates,
and found that $p$-wave Fermi condensates behave very different from
$s$-wave Fermi condensates in the following aspects.

First, the matter-wave interference and the Josephson effect of two $p$-wave Fermi condensates
has an angular effect due to the vector nature of the order parameter.
When the dot product of the vector order parameters of the left and right
condensates reaches its maximum, the interference pattern is most visible and
Josephson current is the largest. However, when
the dot product of vector order parameters is zero, the interference pattern
disappears and the Josephson current vanishes.
This effect is absent in the BEC limit of $s$-wave Fermi superfluids,
as well as in scalar Bose systems. It was also proposed that the relative orientation
of the order parameters of two $p$-wave condensates can be controlled
by applying a constant plus a gradient magnetic field which makes clouds cross
different Feshbach resonances and prepares the vector order parameters in
orthogonal or non-orthogonal configurations. These general considerations
were then applied to the specific case of $^6$Li, where Feshbach resonances are
split into $\vert 11 \rangle$, $\vert 12 \rangle + \vert 21 \rangle$, and $\vert 22 \rangle$,
depending on the different hyperfine states of the colliding atoms.

Second, we showed that anisotropic $p$-wave interactions
lead to anisotropic effective masses for a given orbital symmetry,
as unitary is approached from the BEC regime.
Furthermore, for cigar-shaped clouds with axial symmetry, we found that the cloud expansion is
anisotropic in the radial plane and expands more rapidly along the direction
of smaller effective mass due to the anisotropy of $p$-wave interactions.
We emphasized that this anisotropic expansion is a result of the anisotropic
$p$-wave interactions and occurs even in a completely isotropic trap, in sharp
contrast with the standard anisotropy inversion observed during expansion due to
anisotropic cloud confinements.
In addition, we would like to stress that the anisotropic expansion should occur
not only for $p$-wave, but also for any higher angular momenta ($d$-wave, $f$-wave, etc...)
Feshbach resonances. Lastly, we pointed out that
the orbital symmetry of the order parameter for $p$-wave condensates can be
directly probed through cloud expansions and that a potential candidate for such
experiments is $^{40}$K, where $p$-wave Feshbach resonances are split depending
on the internal angular momentum states $(\ell = 1, m_{\ell} = 0)$ and
$(\ell = 1, m_{\ell} = \pm 1 )$.

We would like to thank NSF for support (Grant No. DMR-0304380).

%%%%%%%%%%%%%%%%%%%%%%%%
%%%%%%%%%%%%%%%%%%%%%%%%

%\begin{references}

%\end{references}

\begin{thebibliography}{99}

\bibitem{andrews-97}
M. R. Andrews, C. G. Townsend, H. -J. Miesner, D. S. Durfee, D. M. Kurn, and W. Ketterle,
Science {\bf 275}, 637 (1997).

\bibitem{shin-04}
Y. Shin, M. Saba, T. A. Pasquini, W. Ketterle, E. D. Pritchard, and A. E. Leanhardt,
Phys. Rev. Lett. {\bf 92}, 050405 (2004).

\bibitem{schumm-05}
T. Schumm, S. Hofferberth, L. M. Andersson, S. Wildermuth, S. Groth,
I. Bar-Joseph, J. Schmiedmayer, and P. Kruger,
Nature Physics {\bf 1}, 57 (2005).

\bibitem{bloch-05}
S. Folling, F. Gerbier, A. Widera, O. Mandel, T. Gericke, and I. Bloch,
Nature {\bf 434}, 481 (2005).

\bibitem{greiner-03}
M. Greiner, C. A. Regal, and D. S. Jin,
Nature (London) {\bf 426}, 537 (2003).

\bibitem{zwierlein-03}
M. W. Zwierlein, C. A. Stan, C. H. Schunck, S. M. F. Raupach,
S. Gupta, Z. Hadzibabic, and W. Ketterle,
Phys. Rev. Lett. {\bf 91}, 250401 (2003).

\bibitem{bartenstein-04}
M. Bartenstein, A. Altmeyer, S. Riedl, S. Jochim, C. Chin,
J. Hecker Denschlag, and R. Grimm,
Phys. Rev. Lett. {\bf 92}, 120401 (2004).

\bibitem{bourdel-04}
T. Bourdel, L. Khaykovich, J. Cubizolles, J. Zhang,
F. Chevy, M. Teichmann, L. Tarruell, S. J. J. M. F. Kokkelmans, and C. Salomon,
Phys. Rev. Lett. {\bf 93}, 050401 (2004).

\bibitem{thomas-04}
J. Kinast, S. L. Hemmer, M. E. Gehm, A. Turlapov and J. E. Thomas,
Phys. Rev. Lett {\bf 92}, 150402 (2004).

\bibitem{partridge-05}
G. B. Partridge, K. E. Strecker, R. I. Kamar, M. W. Jack, and R. G. Hulet,
Phys. Rev. Lett. {\bf 95}, 020404 (2005).

\bibitem{sademelo-93}
C. A. R. S{\'a} de Melo, M. Randeria, and J. R. Engelbrecht,
Phys. Rev. Lett. {\bf 71}, 3202 (1993).

\bibitem{iskin-06a}
M. Iskin and C. A. R. S{\'a} de Melo, Phys. Rev. Lett. {\bf 96}, 040402 (2006).

\bibitem{regal-03b}
C. A. Regal, C. Ticknor, J. L. Bohn, and D. S. Jin,
Phys. Rev. Lett. {\bf 90}, 053201 (2003).

\bibitem{ticknor-04}
C. Ticknor, C. A. Regal, D. S. Jin, and J. L. Bohn,
Phys. Rev. A {\bf 69}, 042712 (2004).

\bibitem{zhang-04}
J. Zhang, E. G. M. van Kempen, T. Bourdel, L. Khaykovich, J. Cubizolles,
F. Chevy, M. Teichmann, L. Tarruell, S. J. J. M. F. Kokkelmans, and C. Salomon,
Phys. Rev. A {\bf 70}, 030702(R) (2004).

\bibitem{schunck-05}
C. H. Schunck, M. W. Zwierlein, C. A. Stan, S. M. F. Raupach,
W. Ketterle, A. Simoni, E. Tiesinga, C. J. Williams, and P. S. Julienne,
Phys. Rev. A {\bf 71}, 045601 (2005).

\bibitem{gunter-05}
K. Gunter, T. Stoferle, H. Moritz, M. Kohl, and T. Esslinger,
Phys. Rev. Lett. {\bf 95}, 230401 (2005).

\bibitem{botelho-04}
S. S. Botelho and C. A. R. S{\'a} de Melo,
cond-mat/0409357.

\bibitem{botelho-05}
S. S. Botelho and C. A. R. S{\'a} de Melo,
J. Low Temp. Phys. {\bf 140}, 409 (2005).

\bibitem{gurarie-05}
V. Gurarie, L. Radzihovsky, and A. V. Andreev,
Phys. Rev. Lett. {\bf 94}, 230403 (2005).

\bibitem{yip-05}
C.-H. Cheng and S.-K. Yip,
Phys. Rev. Lett. {\bf 95}, 070404 (2005).

\bibitem{ohashi-05}
Y. Ohashi,
Phys. Rev. Lett. {\bf 94}, 050403 (2005).

\bibitem{ho-05}
T.-L. Ho and R. B. Diener,
Phys. Rev. Lett. {\bf 94}, 090402 (2005).

\bibitem{iskin-06b}
M. Iskin and C. A. R. S{\'a} de Melo,
Phys. Rev. A {\bf 74}, 013608 (2006).

\bibitem{zhang-06}
W. Zhang and C. A. R. S{\'a} de Melo,
cond-mat/0603601 (2006).

\bibitem{tokatly-04}
I. V. Tokatly, Phys. Rev. A {\bf 70}, 043601 (2004).

\bibitem{kagan-96}
Y. Kagan, E. L. Surkov, and G. V. Shlyapnikov,
Phys. Rev. A {\bf 54}, R1753 (1996).

\bibitem{castin-96}
Y. Castin and R. Dum, Phys. Rev. Lett. {\bf 77}, 5315 (1996).

\bibitem{vaccarella-03a}
C. D. Vaccarella, R. D. Duncan, and C. A. R. {S\'a} de Melo,
cond-mat/0302080 (2003).

\bibitem{vaccarella-03b}
C. D. Vaccarella, R. D. Duncan, and C. A. R. {S\'a} de Melo,
Physica C {\bf 391}, 89 (2003).

\bibitem{zwierlein-04-prl}
M.W. Zwierlein, C. A. Stan, C. H. Schunck, S.M. F. Raupach, A. J. Kerman, and W. Ketterle,
Phys. Rev. Lett. {\bf 92}, 120403 (2004).

\bibitem{chin-06}
J. K. Chin, D. E. Miller, Y. Liu, C. Stan, W. Setiawan, C. Sanner, K. Xu, and  W. Ketterle,
Nature {\bf 443}, 961 (2006).

\bibitem{menotti-02}
C. Menotti, P. Pedri, and S. Stringari,
Phys. Rev. Lett. {\bf 89}, 250402 (2002).

\bibitem{naraschewski-96}
M. Naraschewski, H. Wallis, A. Schenzle, J. I. Cirac, and P. Zoller,
Phys. Rev. A {\bf 54}, 2185 (1996).

\bibitem{schunck-06}
C. H. Schunck, M. W. Zwierlein, A. Schirotzek, and W. Ketterle,
Phys. Rev. Lett. {\bf 98}, 050404 (2007).

\end{thebibliography}
\end{document}